\documentclass{iopconfser}

\usepackage{subcaption}
\usepackage[T1]{fontenc}
\usepackage{float}
\usepackage{latexsym}
\usepackage{amsmath,amstext,amssymb,amsfonts,
amscd,bm,array,multirow,amsbsy,mathrsfs,calc}
\usepackage{t1enc}
\usepackage{indentfirst}
\usepackage{pb-diagram}
\usepackage{graphicx}
\usepackage{enumerate}
\usepackage{color}
\usepackage[all]{xy}
\usepackage{hyperref}
\hypersetup{colorlinks=false,pdfborderstyle={/S/U/W 0}}
\usepackage[usenames,dvipsnames]{xcolor}
\usepackage{url}
\usepackage{upgreek}

\newcommand{\be}{\begin{equation*}}
\newcommand{\ee}{\end{equation*}}
\newcommand{\ben}{\begin{equation}}
\newcommand{\een}{\end{equation}}
\newcommand{\beqa}{\begin{eqnarray*}}
\newcommand{\eeqa}{\end{eqnarray*}}
\newcommand{\beqan}{\begin{eqnarray}}
\newcommand{\eeqan}{\end{eqnarray}}
\newcommand{\nn}{\nonumber}

\def\i{\mathbf{i}}

\def\C{\mathbb{C}}
\def\R{\mathbb{R}}

\def\End{\mathrm{End}}
\def\Hess{\mathrm{Hess}}

\def\Crit{\mathrm{Crit}}

\def\pd{\partial}

\def\dd{\mathrm{d}}

\newcommand{\Tr}{\mathrm{Tr}}

\newcommand{\sign}{\mathrm{sign}}

\def\cC{\mathcal{C}}
\def\cD{\mathcal{D}}

\def\cF{\mathcal{F}}
\def\cG{\mathcal{G}}
\def\cH{\mathcal{H}}

\def\cM{\mathcal{M}}

\def\cS{\mathcal{S}}

\def\cO{\mathcal{O}}

\def\rS{\mathrm{S}}

\newcommand{\eqdef}{\stackrel{{\rm def.}}{=}}

\def\grad{\mathrm{grad}}
\def\Sym{\mathrm{Sym}}
\def\End{\mathrm{End}}

\def\Re{\mathrm{Re}}
\def\Im{\mathrm{Im}}

\def\vol{\mathrm{vol}}

\begin{document}

\title{On consistency conditions for strong SRRT inflation in two-field cosmological models}

\author{Elena-Mirela Babalic, Calin-Iuliu Lazaroiu}

\affil{Department of Theoretical Physics, National Institute for R\&{D} in Physics and Nuclear Engineering, Str. Reactorului 30, PO Box MG-6, 077125 Magurele, Ilfov, Romania}

\email{mbabalic@theory.nipne.ro, lcalin@theory.nipne.ro}

\begin{abstract}
We discuss the strong version of the consistency conditions for SRRT
inflation in general two-field cosmological models. In the ``fiducial'' case,
this condition is a geometric PDE which relates the scalar field
metric and scalar potential of such models. When supplemented by appropriate
boundary conditions, this equation determines the scalar field metric in terms of
the scalar potential or the other way around, thereby selecting “fiducial”
models for strong SRRT inflation. When the scalar potential is given,
the equation can be simplified by fixing the conformal class of the
scalar field metric, in which case it locally becomes an equation for
the conformal factor of that metric when written in isothermal
coordinates.  We analyze this equation with standard methods of PDE
theory, discuss its quasilinearization near a non-degenerate critical
point of the scalar potential and extract natural asymptotic
conditions for its solutions near such points.
\end{abstract}

\section{Introduction}

Inflation provides the dominant theoretical framework for
understanding physics of the very early Universe. It successfully
accounts for the observed homogeneity, isotropy, and flatness of the
cosmos, while also offering a mechanism for generation of
primordial perturbations that seed the large-scale structure. This
paradigm remains in excellent agreement with current cosmological
observations, including the latest measurements of the cosmic
microwave background (CMB) anisotropies and polarization from Planck
and WMAP, as well as large-scale structure surveys such as DESI and
Euclid.

While the simplest models of inflation involve a single scalar field
(the inflaton) minimally coupled to gravity, more general scenarios
with multiple scalar fields arise naturally in fundamental high-energy
theories, such as supergravity and string theory. In those contexts,
the inflaton sector typically includes a rich moduli space of scalar
fields, reflecting the geometry of the compactification
manifold. Consequently, multifield inflationary models serve not only
as phenomenological extensions of single-field inflation but also as
probes of ultraviolet (UV) complete theories of nature. Their study
thus plays a central role in the program of testing string theory and
related frameworks through precision cosmological data.  Furthermore, 
recent arguments suggest \cite{S2,S3,S4} that such models may be
preferred in all consistent theories of quantum gravity.

In a multifield model of this type, the classical action describes a set
of scalar fields with canonical couplings to gravity, where the scalar
fields correspond to coordinates on a (connected but generally
non-compact) differentiable manifold $\cM$ called the target
space. The dimension of $\cM$ gives the number of scalar degrees of
freedom. The dynamics is governed by a Riemannian metric $\cG$ on
$\cM$ (which determines the kinetic term of the scalar field
lagrangian) and by a real-valued scalar potential $V:\cM \to \R$ which
encodes the interactions among the fields. In general, the kinetic
term metric $\cG$ is not flat, the scalar potential $V$ is
non-constant and the scalar manifold $\cM$ need not be
contractible. Such models generally have rich dynamics, including
curved trajectories in field space, non-adiabatic perturbations and
distinctive non-Gaussian signatures. A nontrivial topology of the
scalar field space $\cM$ has important consequences for dynamics, as
already discussed in \cite{genalpha, elem, modular} and in more
generality in \cite{ren, grad, natobs}.

The simplest nontrivial realization of this framework involves two
scalar fields, a setup that already captures most of the qualitatively
new features of multifield inflation. Compared with the single-field
case, two-field models exhibit significantly richer dynamical
structure, including curved trajectories, entropic (isocurvature)
modes and mode couplings between curvature and isocurvature
perturbations. These effects can lead to observable imprints on the
CMB and large-scale structure formation, such as scale-dependent
non-Gaussianities, correlated isocurvature modes and deviations from
the standard consistency relations.

Recent studies focused on understanding the dynamics of two-field
cosmological models in specific dynamical regimes, notably the
slow-roll slow-turn (SRST) and slow-roll rapid-turn (SRRT)
regimes. These regimes provide useful approximations for
characterizing inflationary trajectories and cosmological
perturbations. The well-known SRST regime generalizes the usual
slow-roll approximation to the multifield context, assuming both small
field accelerations and small turning rates. In contrast, the SRRT
regime allows for significant curvature in field space trajectories
(large turning rate) while maintaining slow evolution of the
background fields. 
Cosmological trajectories with sustained “rapid” turn and slow roll 
are of significant phenomenological interest \cite{RT6,RT8}.

Two variants of the SRRT regime were considered until now in the
literature, namely the so-called {\em strong} \cite{cons, consproc}
and {\em weak} \cite{LiliaCons, Achucarro} SRRT regimes. They are
distinguished by the precise conditions imposed on relative magnitude
of the turning rate and slow-roll parameters. The strong SRRT regime
is particularly interesting because it can sustain prolonged inflation
even in steep potentials, with the turning motion providing an
effective stabilization mechanism. Lazaroiu and Anguelova have
analyzed the dynamical consistency conditions \cite{cons,consproc}
required for the strong SRRT regime to hold, showing that these can
serve as robust selection criteria for constructing physically viable
two-field inflationary models. Their work provides a geometrically
and dynamically natural framework for classifying and constraining multifield
inflationary dynamics, contributing to the broader goal of connecting
high-energy theory with cosmological observations.

Ongoing research continues to refine our understanding of these
regimes, including the role of field-space curvature, multifield
attractors, and reheating dynamics, as well as the connection between
multifield dynamics and observable quantities such as the spectral
tilt, tensor-to-scalar ratio, and non-Gaussianity parameters. As
future missions such as LiteBIRD and CMB-S4 improve observational
sensitivity to primordial signatures, multifield inflation will remain
a crucial testing ground for fundamental physics beyond the Standard
Model.
 
In this paper, which summarizes some results of \cite{axion1},
we discuss solutions of the PDE resulting from the
strong SRRT consistency conditions for general two-field models and
its solutions. By fixing the potential $V$ and the conformal class of
the scalar field metric $\cG$, we seek solutions for the halved
conformal exponent of the metric.

\section{Two-field cosmological models - definition and dynamics}

In the physics literature, the precise definition of two-field
cosmological models is often unclear since the topology of the scalar
field space is not clearly specified. In this paper, we use the
precise mathematical description introduced in \cite{genalpha}, which
allows for nontrivial topology of the space where the scalar fields of
the model take values. Dynamical effects related to non-trivial field
space topology (i.e. non-contractible field spaces) were already
discussed in loc. cit. and explored further in \cite{elem,
  modular}. We stress that there is no apriori physics reason to
assume the scalar field space of a multifield cosmological model to be
topologically trivial.

As proposed in \cite{genalpha} and further discussed in \cite{ren}, a
two-dimensional cosmological model with oriented scalar field
space can be described mathematically as follows.

\

\noindent {\bf Definition 1.} A two-dimensional {\em oriented scalar
  triple} is an ordered system $(\cM,\cG,V)$, where $(\cM,\cG)$ is a
connected, oriented and borderless Riemann surface (called {\em scalar
  manifold}) and $V\in \cC^\infty(\cM,\R)$ is a smooth real-valued
function defined on $\cM$ which is called {\em scalar potential}. The
Riemannian metric $\cG$ on surface $\cM$ is called {\em scalar field
  metric}.

\

\noindent To ensure conservation of energy in our models, we will
assume throughout that the Riemannian manifold $(\cM,\cG)$ is
complete. For simplicity, we also assume that the scalar potential is
strictly positive (i.e. $V>0$ on $\cM$) in order to avoid certain
technical problems (this second condition can be relaxed).

Each two-dimensional oriented scalar triple $(\cM,\cG,V)$ defines a
model of gravity coupled to scalar fields on a spacetime of
topology\footnote{Of course, we consider the {\em standard} topology
on $\R^4$.}  $\R^4$ through the action:
\ben
\label{Action}
\cS_{\cM,\cG,V}[g,\varphi] = \int_{\R^4} \dd^4 x
\,\sqrt{|g|} \left[ \frac{M^2}{2}R(g) - \frac{1}{2}
  \Tr_g\varphi^\ast(\cG) - V\circ \varphi \right]~,
\een
where $g$ is a Lorentzian metric of ``mostly plus'' signature defined
on $\R^4$ and $\varphi:\R^4\rightarrow \cM$ is a map from $\R^4$ into
the surface $\cM$. Here $M$ is the reduced Planck mass while $R(g)$ is
the scalar curvature of the Lorentzian metric $g$ -- hence the first
term in $\cS_{\cM,\cG,V}$ is the Einstein-Hilbert action for
$g$. Choosing local coordinates on $\cM$, the second term in the
Lagrangian above expands as:
\be
[\Tr_g\varphi^\ast(\cG)](x)= g^{\mu\nu}(x)\cG_{ij}(\varphi(x))\partial_\mu\varphi^i(x)\partial_\nu\varphi^j(x)~~,~~\mu,\nu\in\{0,..,3\}~,~i,j\in\{1,2\}~,
\ee
which is the well-known local expression for the standard kinetic term
of the nonlinear sigma model defined on the Lorentzian manifold
$(\R^4,g)$ and with target $(\cM,\cG)$. It is convenient for what
follows to use the {\em rescaled Planck mass}:
\be
M_0\eqdef \sqrt{\frac{2}{3}} M~.
\ee
Taking the spacetine metric $g$ to describe a spatially flat FLRW universe of scale factor 
$a(t)>0$, the squared line element has the following well-known expression:
\be
\dd s_g^2 := - \dd t^2 + a^2(t) \dd \vec{x}^2~,~\mathrm{where}~t\eqdef x^0~\mathrm{and}~ \vec{x}=(x^1, x^2,x^3)~.
\ee
We also take the scalar fields $\varphi:\R^4\to\cM$ to depend only on
the cosmological time $t$, i.e. $\varphi=\varphi(t)$.

Besides the well-known {\em Hubble parameter} $H(t)\eqdef
\frac{\dot{a}(t)}{a(t)}$, we consider the {\em rescaled Hubble
  function}:
\be
\cH:T\cM\rightarrow \R_{>0}~~,~~\cH(u)\eqdef \frac{1}{M_0}\sqrt{||u||^2+2V(\pi(u))}~,~\forall u\in T\cM~~,
\ee
where $\pi:T\cM\rightarrow \cM$ is the bundle projection. Here
$||\cdot||: T\cM\rightarrow \R_{\geq 0}$ is the norm function induced
by $g$ on the total space of the tangent bundle to $\cM$. This
function vanishes precisely on the image of the zero section of
$T\cM$. The rescaled Hubble function is strictly positive and smooth
on $T\cM$ due to our assumption that $V$ is strictly positive on
$\cM$.

With these notations and when $H(t)>0$, the equations of motion of the
model subject to the ansatze above are equivalent with the {\em
  cosmological equation}:
\ben
\label{eom}
\nabla_t \dot{\varphi}(t)+\frac{1}{M_0}\cH(\dot{\varphi}(t)) \dot{\varphi}(t)+ (\grad V)(\varphi(t))=0~~
\een
(here $\nabla_t\eqdef \nabla_{\dot{\varphi}(t)}$, while $\grad V\in
\Gamma(T\cM)$ is the gradient vector field of $V$ relative to the
Riemannian metric $\cG$) together with the {\em Hubble condition}:
\be
H(t)=\frac{1}{3M_0}\cH(\dot{\varphi}(t))~.
\ee
The latter determines the Hubble parameter $H$ as a function of the
cosmological time $t\in I$ given a solution $\varphi:I\rightarrow \cM$
of the cosmological equation, where $I\subset \R$ is a cosmological
time interval which is not reduced to a point. Notice that a maximal
solution $\varphi$ of the cosmological equation need not be defined
for all cosmological times, in which case one must replace $\R^4$ in
the action \eqref{Action} by $I_{\mathrm{max}}\times \R^3$, where
$I_{\mathrm{max}}$ is the maximal interval of definition of $\varphi$
(which is necessarily an open interval and hence diffeomorphic with
$\R$). For simplicity we shall take $M=1$ from now on, which amounts
to setting $M_0=\sqrt{\frac{2}{3}}$. This corresponds to working in
natural units.

For reader's convenience, we mention that the local coordinate forms
of the objects appearing in \eqref{eom} are:
\beqa
&&\nabla_t\varphi^i(t)= \ddot{\varphi}^i(t) + \Gamma^i_{jk}(\varphi(t))\dot{\varphi}^j(t) \dot{\varphi}^k(t) ~~,~\nn\\
&&||\dot{\varphi}(t)||^2 = \cG_{ij}(\varphi(t))\dot{\varphi}^i(t)\dot{\varphi}^j(t) ~~,
\\
&&\grad V=\cG^{ij}(\pd_j V)\pd_i~~,~~\pd_i =\frac{\pd }{\pd \varphi^i}~.\nn
\eeqa
The solutions $\varphi:I\rightarrow \cM$ of the cosmological equation
\eqref{eom} are called {\em cosmological curves}. The cosmological
equation defines a dissipative geometric dynamical system on the
four-dimensional total space of the tangent bundle $T\cM$ (see \cite{ren}).

\section{Slow roll and rapid turn parameters and regimes}
Let $(T,N)$ be the positive\footnote{I.e. positively-oriented relative
to the given orientation of $\cM$.} Frenet frame of a cosmological
curve $\varphi:I\rightarrow \cM$:
\be
T(t)\eqdef \frac{\dot{\varphi}(t)}{||\dot{\varphi}(t)||}~~,~~N(t)=-J T(t)~~,
\ee
where $J\in \End(T\cM)$ is the complex structure determined on $\cM$ by the conformal class of $\cG$:
\be
\omega(X,Y)=\cG(JX,Y)~~,~~X,Y\in T\cM~.
\ee
Here $\omega\eqdef \vol_\cG\in \Omega^2(\cM)$ is the volume form
defined by $\cG$ relative to the given orientation of $\cM$. Let
$\sigma$ be an increasing proper legth parameter for $\varphi$:
\be
\dd \sigma=||\dot{\varphi}(t)||\dd t~~, ~~\mathrm{i.e.}~~~\dot\sigma=||\dot{\varphi}(t)||=\sqrt{\cG_{ij}(\varphi(t))\dot{\varphi}^i(t) \dot{\varphi}^j(t)}~.
\ee
Projecting the cosmological equation \eqref{eom} along $T$ and $N$ gives the well-known {\em adiabatic and entropic equations}:
\be
\ddot{\sigma}+\cH (\sigma,\dot{\sigma}) \dot{\sigma}+V_T(\sigma)=0~~,~~\Omega(\sigma)=\frac{V_N(\sigma)}{\dot{\sigma}}~,
\ee
where:
\beqa
&&\cH(\sigma,\dot{\sigma})=\sqrt{\dot{\sigma}^2+2 V(\sigma)}~~,\\
&& V_T(\sigma)\eqdef T^i(\sigma)(\pd_iV)(\varphi(\sigma))~~,~~V_N(\sigma)\eqdef N^i(\sigma)(\pd_iV)(\varphi(\sigma)) ~~,\\
&& \Omega \eqdef -\cG(N,\nabla_t T)=N_i\nabla_t T^i~~.
\eeqa
The quantity $\Omega(t)$ is called the {\em signed turn rate} of $\varphi$ at cosmological time $t$.

\

\noindent {\bf Definition 2.}
The {\em opposite relative acceleration vector} of the cosmological curve $\varphi$ is defined through:
\be
\eta(t)\eqdef -\frac{1}{H(\varphi(t),\dot{\varphi}(t))\dot{\sigma}(t)}\nabla_t\dot{\varphi}(t)
\ee

\noindent This vector decomposes as $\eta(t)=\eta_\parallel(t)
T(t)+\eta_\perp(t) N(t)$, where $\eta_\parallel(t)$ and
$\eta_\perp(t)$ are real numbers.

\

\noindent {\bf Definition 3.}
The following functions of $t$ associated to the cosmological curve $\varphi$ can be defined:\\
- The first, second and third {\em slow roll parameters}:
\be
\varepsilon\eqdef-\frac{\dot{H}}{H^2}~~,~~\eta_\parallel\eqdef-\frac{\ddot{\sigma}}{H\dot{\sigma}}~~,~~\xi\eqdef\frac{\dddot{\sigma}}{H^2\dot{\sigma}}~,
\ee
- The first and second {\em turn parameters}: 
\be
\eta_\perp\eqdef \frac{\Omega}{H}~~,~~\nu\eqdef \frac{\dot{\eta_\perp}}{H \eta_\perp}~,
\ee
- The {\em first IR parameter} $\kappa$ and the {\em conservative
  parameter} $c$:
\be
\kappa\eqdef \frac{\dot{\sigma}^2}{2 V}~~,~~c\eqdef \frac{H\dot{\sigma}}{||\dd V||}~.
\ee

\noindent {\bf Definition 4.}
Using the first, second and third slow roll conditions ($\epsilon\ll
1$, $|\eta_\parallel|\ll 1$ and $|\xi|\ll 1$), the following regimes
can be identified:\\
- The {\em first order slow roll regime} - which holds when $\epsilon\ll 1$,\\
- The {\em second order slow roll regime} - when $\epsilon\ll 1$ and $|\eta_\parallel|\ll 1$,\\
- The {\em third order slow roll regime} - when $\epsilon\ll 1$, $|\eta_\parallel|\ll 1$ and $|\xi|\ll  1$.

\vspace{2mm}

\noindent {\em Remark 1.} There exist various variants of the rapid turn regime, such as:\\
- The {\em weak rapid turn regime}, defined by: $\eta_\perp^2\gg \max(\epsilon,\eta_\parallel,\xi)$,\\
- The {\em strong rapid turn regime}, defined by: $\eta_\perp^2\gg 1$,\\
- The {\em sustained strong rapid turn regime}, defined by: $\eta_\perp^2\gg 1$ and $|\nu|\ll 1$.

\vspace{2mm}

\noindent {\bf Definition 5.}
We say that $\varphi$ satisfies the {\em strong SRRT conditions}
(third order slow-roll plus sustaind strong rapid-turn) at
cosmological time $t$ iff the following five conditions are satisfied
{\em simultaneously}:
\be
\epsilon(t),~ |\eta_\parallel(t)|,~|\xi(t)|,~|\nu(t)|\ll 1 ~,~\eta_\perp(t)^2\gg 1 ~.
\ee

The following result was established in \cite{cons}.

\noindent {\bf Proposition 1.}
When $|\eta_\parallel|\ll 1$, the strong rapid turn condition
$\eta_\perp^2\gg 1$ is equivalent with $c^2\ll 1$.

\

\noindent {\bf Definition 6.} Let $\cM_0\eqdef \{m\in \cM~\vert~(\dd
V)(m)\neq 0\}$ be the complement of the critical locus of $V$. The
{\em adapted frame} of the scalar triple $(\cM,\cG,V)$ is the
positively-oriented orthonormal frame $(n,\tau)$ of the open
submanifold $\cM_0$ of $\cM$ defined by the following vector fields:
\be
n\eqdef \frac{\grad V}{||\grad V||}~~,~~\tau=-J n=-\frac{\grad_J V}{||\grad V||}~~.
\ee

\noindent We assume from now on that $V$ is not constant on $\cM$, which insures that $\cM_0$ is non-empty. 

\vspace{2mm}

\noindent {\em Remark 2.}
In positively-oriented local coordinates $(U,x^1,x^2)$ on $\cM$, we have:
\be
\grad_JV\eqdef J\grad V=_U \pd^i V J\pd_i=\epsilon_j^{\,\, i} \pd^j V \pd_i=-\epsilon^{ij} \pd_j V \pd_i~~.
\ee

\noindent {\bf Definition 7.}
The {\em characteristic angle} $\theta\in [-\pi,\pi]$ of $\varphi$ is
  the angle of rotation from the adapted frame $(n,\tau)$ 
  to the Frenet frame $(T,N)$ of $\varphi$:
\be
T=n\cos\theta +\tau \sin\theta~~,~~N=-n\sin\theta+\tau\cos\theta~~.
\ee
The components of the relative acceleration vector $\eta$ take the
following form when written in terms of the characteristic angle
$\theta$:
\be
\eta_\parallel=3+\frac{\cos\theta}{c}~~,~~\eta_\perp=-\frac{\sin\theta}{c}~~.
\ee
The adiabatic and etropic equation take the following form in the Frenet frame:
\be 
\frac{V_{TT}}{3 H^2}=\frac{\Omega^2}{3 H^2}+\varepsilon+\eta_\parallel-\frac{\xi}{3} ~~,~~~\frac{V_{TN}}{H^2}=\frac{\Omega}{H}\left(3-\varepsilon-2\eta_\parallel+\nu\right)~.
\ee

\vspace{1mm}
Here and below, we use notation $V_{XY}\eqdef \Hess(V)(X,Y)$ for any
vector fields $X,Y\in T\cM$, where $\Hess(V)\eqdef \nabla \dd V$ is
the Riemannian Hessian tensor of $V$.
  
Suppose that $\varepsilon\ll 1$, $|\eta_\parallel|\ll 1$, $|\xi|\ll
1$, $|\nu|\ll 1$ are satisfied. Then we have $\cos\theta\approx -3
c,~\sin\theta\approx s\sqrt{1-9c^2}$ (where $s\eqdef \sign(\sin\theta)
\in \{-1,0,1\}$) and:
\beqa
&& V_{TT}\approx 9 c^2 V_{nn}-6 s c\sqrt{1-9c^2} V_{n\tau} +(1-9c^2)V_{\tau\tau}\nn\\
&& V_{TN}\approx -3 s c \sqrt{1-9c^2}  (V_{\tau\tau}-V_{nn})-(1-18 c^2) V_{n\tau}~~.
\eeqa
After some tedious manipulations, it was found in \cite{cons} that
these equations admit a solution $c$ with $c^2\ll 1$
(i.e. $\eta_\perp^2\gg1$) iff the following approximate condition
holds:
\be
V_{n\tau}^2 V_{\tau\tau}\approx 3 V V_{nn}^2~.
\ee
This result can be stated more precisely as follows.

\vspace{1mm}

\noindent {\bf Theorem 1.} [Anguelova \& Lazaroiu, 2022]. In the adapted
frame $(n,\tau)$ of $(\cM,\cG,V)$, a cosmological curve
$\varphi:I\rightarrow \cM_0$ whose image is contained in the
noncritical submanifold $\cM_0$ satisfies the strong SRRT conditions
(i.e. the sustained strong rapid turn conditions: $\eta_\perp^2\gg1$,
$|\nu|\ll 1$, with third order slow roll conditions: $\varepsilon\ll
1$, $|\eta_\parallel|\ll 1$, $|\xi|\ll 1$) at cosmological time $t\in
I$ iff the following approximate condition is satisfied at the point
$m=\varphi(t)$ of $\cM_0$:
\be
V_{n\tau}^2 V_{\tau\tau}\approx 3 V V_{nn}^2~~.
\ee

\section{The strong SRRT equation}

It is conceptually convenient to consider the strict form of the approximate condition
discussed in the previous section.

\

\noindent  {\bf Definition 8.}
The {\em strong SRRT equation} is the following condition which
constrains the target space metric $\cG$ and scalar potential $V$ on
the noncritical submanifold $\cM_0$ of the scalar triple $(\cM,\cG,V)$:
\ben
\label{s-SRRT}
V_{n\tau}^2 V_{\tau\tau}=3 V V_{nn}^2~,
\een
where:
\be
 V_{XY}\eqdef \Hess(V)(X,Y) ~~,~~\Hess V=\nabla\dd V~~,~~\nabla=\mathrm{Levi-Civita~connection~on}~\cM~.
\ee

\vspace{2mm}

\noindent {\em Remark 3.} Similarly, one derives the following expression for
the {\em weak SRRT equation} (see \cite{LiliaCons}):
\ben
\label{w-SRRT}
V_{n\tau}^2(V_{nn} V_{\tau\tau} -V_{n\tau}^2)=3V V_{nn} (V_{n\tau}^2+ V_{nn}^2)~.
\een
The strong SRRT equation amounts to a nonlinear partial differential equation
for the pair $(\cG,V)$ on $\cM_0$. When $\cG$ is fixed, it can be
viewed as a nonlinear second order PDE for $V$.  When $V$ is fixed, it
can be viewed as a nonlinear first order PDE for the metric $\cG$.

\subsection{Viewing SRRT equation as a contact Hamiltonian-Jacobi equation}

\

\noindent Let $S\eqdef \Sym^2(T^\ast\cM)$ be the vector bundle of
symmetric covariant 2-tensors on $\cM$ and $S_+\subset S$ be the fiber
sub-bundle consisting of strictly positive-definite tensors. When $V$ is fixed,
the SRRT equation has the form:
\be
\cF(j^1(\cG))=0~~,
\ee
where $\cF:j^1(S_+)\rightarrow \R$ is a smooth function which depends on $V$. 

Let $L=\det T^\ast \cM=\wedge^2 T^\ast \cM$ be the real determinant
line bundle of $\cM$ and $L_+$ be its open sub-bundle of positive
vectors. Fixing the complex structure $J$ determined by the conformal
class of $\cG$, the map $\cG\rightarrow \omega$ gives an isomorphism
of fiber bundles $S_+\stackrel{\sim}{\rightarrow} L_+$ which extends
to an isomorphism $j^1(S_+)\stackrel{\sim}{\rightarrow} j^1(L_+)$. Use
this to transport $\cF$ to a function $F:=F^J_V:j^1(L_+)\rightarrow
\R$. Then the SRRT equation becomes:
\ben
\label{HJ1}
F(j^1(\omega))=0~~.
\een
This is a contact Hamilton-Jacobi equation for $\omega\in \Gamma(L_+)$
relative to the Cartan contact structure of $j^1(L_+)$. The contact
Hamiltonian $F$ restricts to a cubic polynomial function on the fibers
of the natural projection $j^1(L_+)\rightarrow L_+$.

In local isothermal coordinates $(x^1,x^2)$ on $\cM$ relative to $J$, we have:
\be
\dd s^2_\cG=e^{2\varphi}(\dd x_1^2+\dd x_2^2)
\ee
and one can write this contact Hamilton-Jacobi equation as a nonlinear first
order PDE for the conformal factor $\varphi$, which is cubic in the
partial derivatives $\pd_1\varphi$ and $\pd_2\varphi$. The equation
can be solved locally through the method of characteristics, while the
Cauchy boundary value problem can be approached using the theory of
viscosity solutions.

In local isothermal coordinates, the complex structure $J$ is given on $U$ by the conditions: 
\be
J\pd_1=\pd_2~~,~~J\pd_2=-\pd_1~~,~~\mathrm{i.e.}~~J\pd_i=\varepsilon_{ij}\pd_j~~.
\ee
We have
\be
\omega_{ij}=J_{ij}=\epsilon_{ij}=f\varepsilon_{ij}~,
\ee
where the Levi-Civita tensor of $\cG$ is:
\be
\epsilon_{ij}=\omega(\pd_i,\pd_j)=f\varepsilon_{ij}~~,~~f\eqdef e^{2\phi}~.
\ee
Here $\phi\in \cC^\infty(U)$ is the halved conformal exponent and $\omega$ is the volume form of the Riemannian metric $\cG$ defined on $\cM$: 
\ben
\label{cGomega}
\cG(X,Y)=\omega(X,JY)~~.
\een
Note that:
\be
\cG=e^{2\phi} \cG_0~,~~~\dd s_\cG^2= e^{2\phi}\dd s_0^2~~,~~\omega=e^{2\phi} \omega_0=e^{2\phi}\dd x^1 \wedge \dd x^2=\frac{\i}{2} e^{2\phi} \dd z\wedge \dd \bar{z}~,
\ee
where $\cG_0$ is the flat metric on $U$. The Christoffel symbols of $\cG$ are given by:
\ben
\label{ChristoffelIsothermal}
\Gamma_{ij}^k=\delta_i^k \pd_j \phi +\delta_j^k \pd_i \phi-\delta_{ij}\pd_k\phi~~,
\een
while its Gaussian curvature (which equals half of the scalar curvature) takes the form:  
\ben
\label{curvature}
K=- e^{-2\phi} \Delta \phi~~,
\een
where $\Delta$ is the Laplacian operator of the Riemannian manifold $(\cM,\cG)$.

\subsection{The frame-free form of the strong SRRT equation}

\

\noindent To write the strong SRRT equation in a more useful form,
recall that the following relations hold in local isothermal coordinates $(U,x^1,x^2)$ on $\cM_0$:
\be
|| \grad  V||= ||\dd V||=_U\sqrt{\pd^iV\pd_iV}~~~, ~~~\Delta V=_U\pd^i\pd_iV-\cG^{ij}\Gamma_{ij}^k\pd_kV~,
\ee
\be
n\!=\!\frac{1}{||\dd V||}\grad V\!=_U\!\frac{1}{||\dd V||} \pd^i V \pd_i ~\quad~,~\quad~\tau\!=\!\frac{1}{||\dd V||}\grad_JV\!=_U\!-\frac{1}{||\dd V||}\epsilon^{ij}\pd_j V \pd_i~,
\ee
Using these relations, we find: 
\ben
\label{VHessOp}
V_{n\tau}=\frac{1}{||\dd V||^2} \cD_1(\cG,V)~~,~~V_{n n}=\frac{1}{||\dd V||^2} \cD_2(\cG,V)~~,~~V_{\tau\tau}=\frac{1}{||\dd V||^2} \cD_3(\cG,V)~,
\een
where: 
\beqan
\label{D1D2D3}
&& \cD_1(\cG,V\!)\!\eqdef\! \Hess_J(V)(\grad V,\grad V)\!=_U\!\!{\tilde V}_{(ij)} \pd^iV\pd^jV~~,\nn\\
&& \cD_2(\cG,V\!)\!\eqdef\! \Hess(V)(\grad V,\grad V)\!=_U\!\! V_{ij}\pd^iV\pd^jV~~,\\
&& \cD_3(\cG,V\!)\!\eqdef\! \Hess(V\!)(J\grad V, \!J\grad V\!)\!=_U\!\! V_{ij}\epsilon^{ik}\epsilon^{jl}\pd_kV\pd_lV\!\!=\!||\dd V||^2 \Delta V\!\!-\!\cD_2(\cG,\!V\!)~~.\nn
\eeqan
With these notations, the SRRT equation takes the following frame-free form:
\ben
\label{s-SRRT-free}
\cD_1(\cG,V)^2 \cD_3(\cG,V)=3V ||\dd V||^2 \cD_2(\cG,V)^2 ~.
\een
One way to simplify the very complicated PDE \eqref{s-SRRT-free} is to
restrict the metric $\cG$ to lie within a fixed conformal class, which
amounts to fixing the complex structure $J$ defined by the conformal
class of $\cG$ relative to the given orientation of $\cM$. Expanding $F$, we have:
 \beqa
&&F= A B^2 -3V e^{2u} A^2+(\Delta_0V-H_0) B^2-2{\tilde H}_0 A B +(6 V e^{2u} H_0+{\tilde H}_0^2) A \nn\\
&&~~~~- 2{\tilde H}_0(\Delta_0V-H_0) B+{\tilde H}_0^2 [(\Delta_0V)-H_0]-3V e^{2 u} H_0^2~,
\eeqa
where we used the notations:
\be 
A\eqdef (\pd_iV)(x) p_i~~,~~B\eqdef -\epsilon_{ij} (\pd_jV)(x) p_i~,
\ee
\be
H_0\eqdef \frac{(\partial_i\partial_j V)(\partial_i V)(\partial_j V)}{||\dd V||_0^2}~, ~~\tilde H_0\eqdef\frac{-(\partial_i\partial_j V)(\partial_i V)\varepsilon_{jk}(\partial_j V)}{||\dd V||_0^2}~,
\ee
with:
\be
||\dd V||_0^2\eqdef (\pd_1V)^2+(\pd_2V)^2~~,~~ \Delta_0 V \eqdef (\pd_1^2 +\pd_2^2) V
\ee
and:
\be
p_1\eqdef \frac{\pd_1V A-\pd_2V B}{(\pd_1V)^2+(\pd_2V)^2}~~,~~p_2\eqdef\frac{\pd_2V A+\pd_1V B}{(\pd_1V)^2+(\pd_2V)^2}~.
\ee
Defining $P_1= A-H_0$ and $~P_2=B-{\tilde H}_0$, the contact Hamiltonian $F$ can be written as:
\ben
\label{FP}
F=  P_2^2[P_1 +  (\Delta_0 V)] - 3 V e^{2 u} P_1^2 ~.
\een

\subsection{The method of characteristics}

\

\noindent The classical method of characteristics can be applied to extract {\em local} solutions of our contact Hamilton-Jacobi equation.

\

\noindent {\bf Theorem 2.}
In isothermal Liouville coordinates $(x^1,x^2,u,p_1,p_2)$, the contact Hamiltonian reduces to the smooth function $F:U_0\times \R^3\rightarrow \R$
given by:
\ben
\label{contactH}
F(x,u,p)\!=\! [B(x)\!-\!\tilde{H}_0(x)]^2 [A(x,p)\!+\!(\Delta_0V)(x)\! -\! H_0(x)]\!-\!3 e^{2 u} V  [A(x,p)\!-\!H_0(x)]^2
\een
and the contact Hamilton-Jacobi equation takes the form:
\ben
\label{HJ}
F(x_1,x_2,u, p_1,p_2)=F(x_1,x_2,\phi, \pd_1\phi,\pd_2\phi)=0~~,
\een
where $u\circ\omega=\phi,~ p_1\circ\omega=\pd_1\phi,~p_2\circ\omega=\pd_2\phi$~.

\

\noindent{\em Remark 4.} The Dirichlet problem can be approached {\em
  globally} using the theory of viscosity solutions, which is related to the the
{\em viscosity perturbation} of the contact Hamilton-Jacobi equation:
\be
F(x_1,x_2,\phi,\pd_1\phi,\pd_2\phi)-v\Delta_0\phi =0~~~~(v=\mathrm{viscosity~parameter})~.
\ee

A {\em characteristic point} of $F$ is a point $(x,u,p)\in U_0\times \R^3$ such that:
\be
F(x,u,p)=F_{p_1}(x,u,p)=F_{p_2}(x,u,p)=0~~.
\ee
The Dirichlet problem for our PDE asks for a solution $\phi$ of the
equation which satisfies the boundary condition:
\ben
\label{Dirichlet}
\phi\circ \gamma=\phi_0~,
\een
where $\gamma:I\rightarrow U_0$ is a non-degenerate smooth curve and
$\phi_0:I\rightarrow \R$ is a smooth function. The characteristic
system of $F=0$ reads ($t$ here is {\em not} the cosmological time):
\beqan
\label{charsys}
&& \frac{\dd x^i}{\dd t}=F_{p_i}(x,u,p)\nn\\
&& \frac{\dd u}{\dd t}=p_i F_{p_i}(x,u,p)\\
&& \frac{\dd p_i}{\dd t}=-F_{x_i}(x,u,p)-p_i F_u(x,u,p)~,\nn
\eeqan
where $F_{x_i}$, $F_u$ and $F_{p_i}$ are the partial derivatives of
$F$ with respect to $x_i$, $u$ and $p_i$. To locally solve the
Dirichlet problem, one searches for a family of solutions
$(x(t,q),u(t,q), p(t,q))$ satisfying the initial conditions:
\ben
\label{incond}
x(0,q)=x_0(q)~~,~~u(0,q)=\phi_0(q)~~,~~p(0,q)=p_0(q)~~(where ~q\in I)~.
\een
From such a family, one then extracts the solution of interest of the
Hamilton-Jacobi equation using the implicit function theorem.

\subsection{Numerical examples}

\noindent The critical points of the scalar potential $V$ play a
crucial role in determining important features of cosmological
dynamics. It is hence natural to study the behavior of the contact
Hamilton-Jacobi equation and its solutions in the vicinity of
non-degenerate critical points of $V$.

We illustrate this with a few solutions of the contact Hamilton-Jacobi
equation for the halved conformal exponent $\phi$.  In the complex
plane $\C$ of complex coordinate $z=x^1+\i x^2$, we take:
\be
V(x_1,x_2)=V_c+\frac{1}{2}(\lambda_1 x_1^2+\lambda_2 x_2^2)~~,
\ee
where $V_c>0$ and $\lambda_1,\lambda_2\in\R_+$ are the principal
values of $\Hess(V)(c)$. This quadratic potential has a single
critical point located at the origin of the $(x_1,x_2)$-plane, which
is an extremum if $\lambda_1\lambda_2>0$ and a saddle point if
$\lambda_1\lambda_2<0$. We consider the Dirichlet problem with
boundary condition:
\be
\phi_0=-\log[R\log(1/R)]
\ee
imposed on a circle of radius $R=\frac{1}{20}$ centered at the origin
of the $(x_1,x_2)$ plane.

In Figure 1 and Figure 2 we exemplify in two cases the potential
contour plot, the 3D plot, the projected characteristic and a
viscosity approximant of the solution of the Dirichlet problem for the
contact HJ equation for fixed $R=\frac{1}{20}$ and $\lambda_2=1$ and
for different values for $V_c$ and $\lambda_1$.

 \begin{figure}[H]
\centering
\begin{minipage}{.43\textwidth}
\centering \!\!\!\!\includegraphics[width=0.85\linewidth]{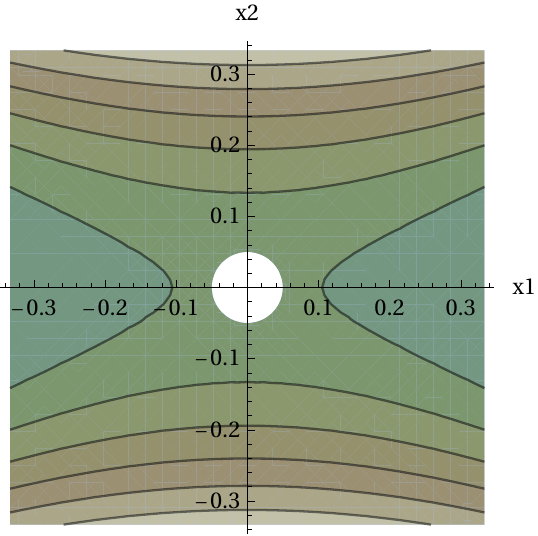}
\subcaption{Contour plot of the potential.}
\end{minipage}
\hfill
\begin{minipage}{.48\textwidth}
\centering \!\!\!\!\!\includegraphics[width=0.95\linewidth]{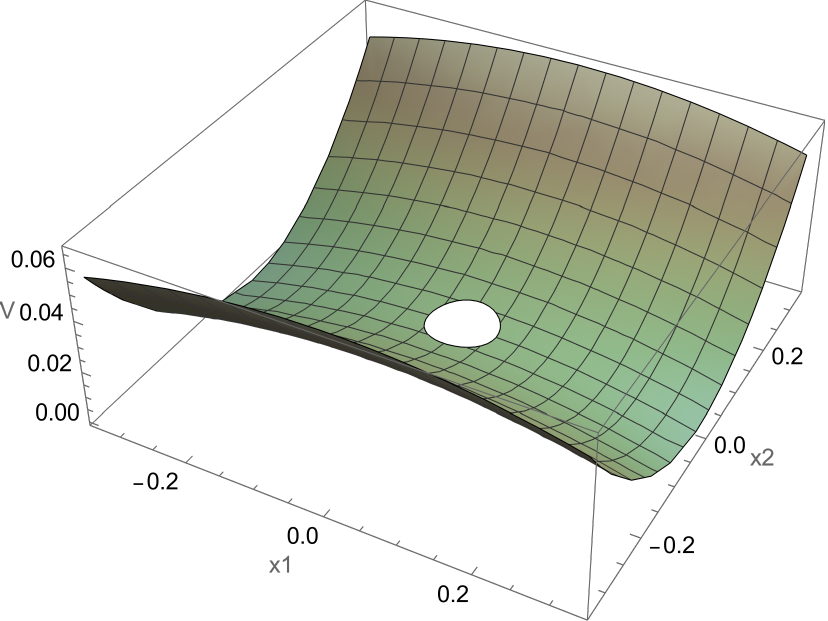}
\subcaption{3D plot of the potential.}
\end{minipage}
\hfill
\centering
\begin{minipage}{.48\textwidth}
\vspace{2em}
\centering ~~ \includegraphics[width=.2\linewidth]{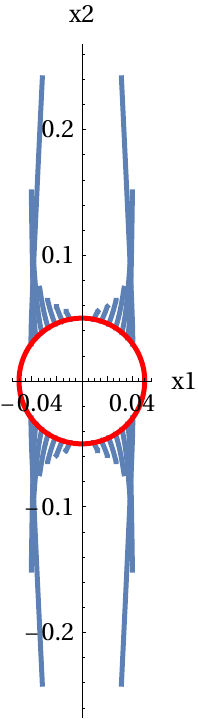}
\subcaption{Some characteristic curves projected on the
  $(x_1,x_2)$-plane. }
\end{minipage}
\hfill
\begin{minipage}{.48\textwidth}
\vspace{2em}
\centering \!\!\!\!\!\!\includegraphics[width=.95\linewidth]{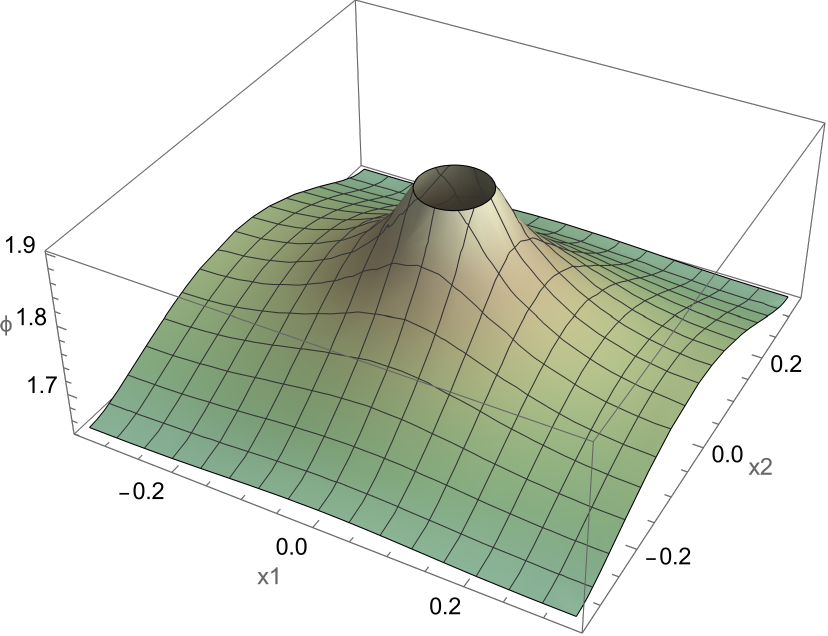}
\subcaption{Solutions of the Dirichlet problem for the viscosity
  perturbation with $v=e^{-7}$.}
\end{minipage}
\hfill
\vspace{2em}
\caption{The potential, projected characteristics and a viscosity
  approximant of the solution of the Dirichlet problem for the contact
  HJ equation for $V_c=1/90$, $\lambda_1=-1/5$,
  $\lambda_2=1$, $R=1/20$.}
\end{figure}

\begin{figure}[H]
\centering
\begin{minipage}{.43\textwidth}
\centering \!\!\includegraphics[width=.8\linewidth]{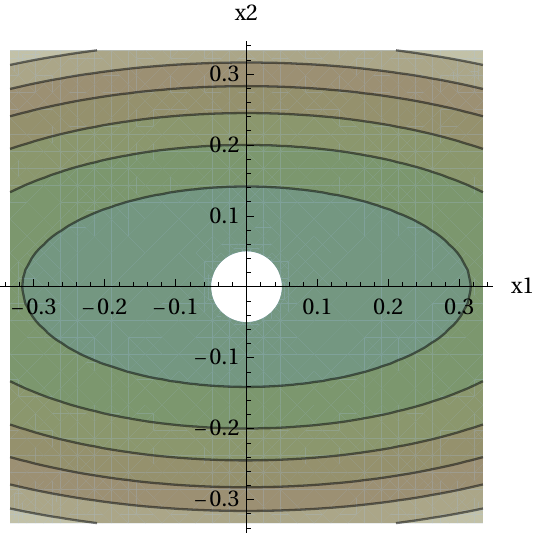}
\subcaption{Contour plot of the potential.}
\end{minipage}
\hfill
\begin{minipage}{.48\textwidth}
\centering \!\!\!\!\!\!\!\!\includegraphics[width=.9\linewidth]{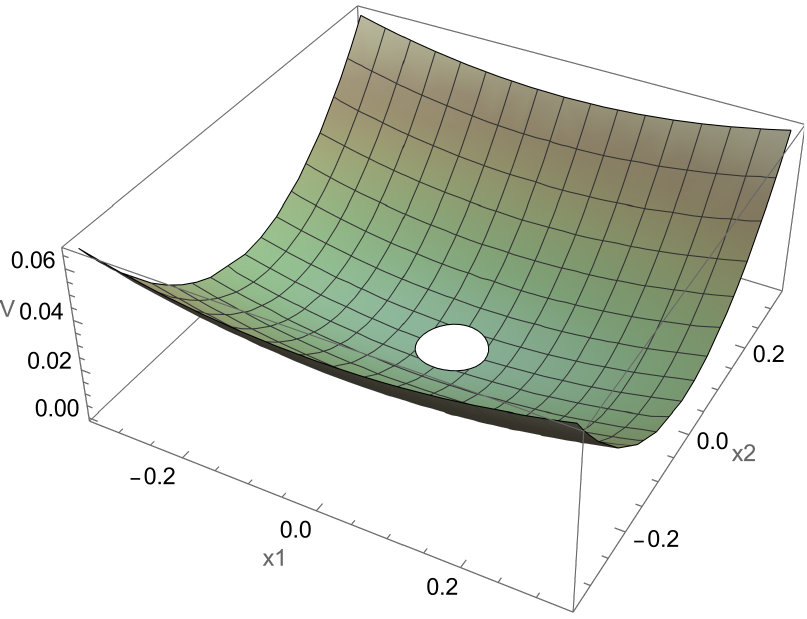}
\subcaption{3D plot of the potential.}
\end{minipage}
\hfill
\\
\centering
\begin{minipage}{.48\textwidth}
\vspace{2em}
\centering \!\!\!\!\!\!\! \includegraphics[width=.7\linewidth]{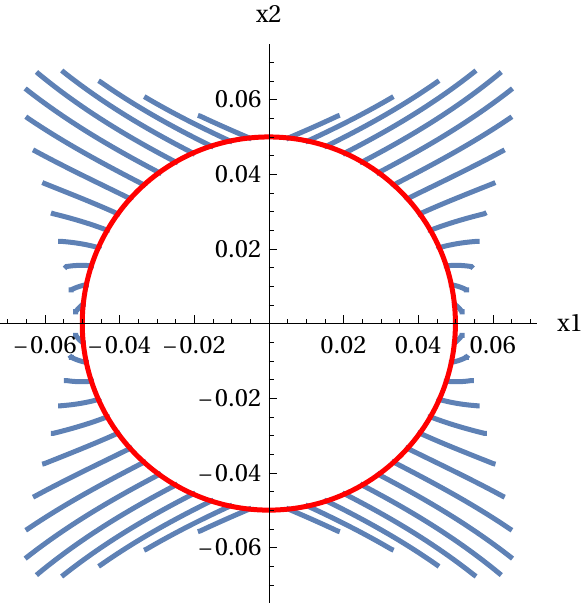}
\subcaption{Some characteristic curves projected \\on the $(x_1,x_2)$-plane.}
\end{minipage}
\hfill
\begin{minipage}{.48\textwidth}
\vspace{2em}
\centering  \includegraphics[width=.9\linewidth]{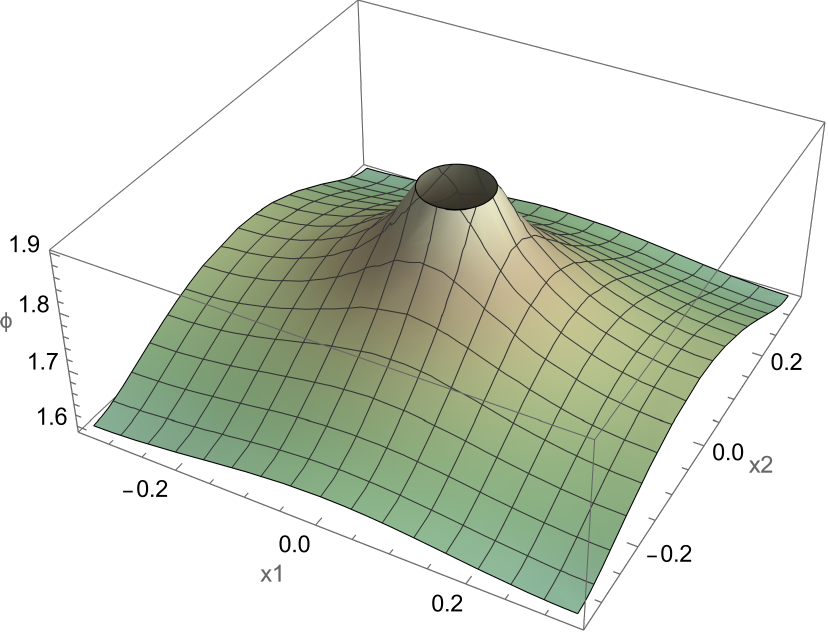}
\subcaption{Solution of the Dirichlet problem for the viscosity perturbation with $v=e^{-7}$.}
\end{minipage}
\hfill
\vspace{1em}
\caption{The potential, projected characteristics and a viscosity
  approximant of the solution of the Dirichlet problem for the contact
  HJ equation for $V_c\!=\!10^{-10}$ and
  $\lambda_1\!=\!1/5$, $\lambda_2\!=\!1$ with $R\!=\!1/20$.}
\label{fig:Nonlinear4}
\end{figure}

\section{The quasilinear approximation near a non-degenerate critical point of the scalar ppotential}
\label{sec:quasilinear}

The critical points of the scalar potential $V$ play a crucial role in
determining important features of cosmological dynamics. In this
section, we study the behavior of the contact Hamilton-Jacobi equation
and its solutions in the vicinity of {\em non-degenerate} critical
points of $V$. We showed in \cite{axion1} that this nonlinear equation can be
approximated by a quasilinear PDE near such a point and studied the
solutions of the latter, which provided asymptotic approximants for certain
solutions of the contact Hamilton-Jacobi equation.

For $c\in \Crit V$ a non-degenerate (hence isolated) critical point
of $V$ and a fixed complex structure $J$ on $\cM$, choosing $(U,z)$ to be a
complex coordinate chart for $(\cM,J)$ centered at $c$ (i.e. $z(c)=0$)
such that the punctured neighborhood $\dot{U}\eqdef U\setminus \{c\}$
is contained in $\cM_0$ and $x^1=\Re z$ and $x^2=\Im z$ are the
corresponding isothermal coordinates defined on $U$, the Riemannian Hessian
of $V$ at $c$ is a non-degenerate bilinear symmetric form defined on
$T_c\cM$ which is independent of the choice of the scalar field metric
$\cG$. In particular, we have:
\be
\Hess(V)(c)=\Hess_0(V)(c)=\frac{1}{2}(\pd_i\pd_j V)(c)\pd_i\otimes \pd_j\vert_{c}\in \Sym^2(T^\ast_c\cM)~~.
\ee
We can assume without loss of generality that the isothermal coordinate system $(U,x)$ was chosen such that:
\be
\pd_i\pd_jV=\lambda_1 \delta_{i1}\delta_{j1}+\lambda_2\delta_{i2}\delta_{j2}~~,
\ee
where $\lambda_1,\lambda_2\!\in \!\R$ are the real eigenvalues of the
Euclidean Hessian operator $\widehat{\Hess}_0(c)\!\in \!\End(T_c\cM)$
at $c$. Non-degeneracy of the Hessian at $c$ implies that $\lambda_1$
and $\lambda_2$ are both nonzero. With this choice of isothermal
coordinates, the Taylor expansion of $V$ at $c$ takes the form:
\be
V(x)=V(c)+\frac{1}{2}(\pd_i\pd_jV)(c)x^i x^j +\cO(||x||_0^3)=V_c+\frac{1}{2}(\lambda_1 x_1^2+\lambda_2 x_2^2) +\cO(||x||_0^3)~~.
\ee

In \cite{axion1}, we showed that the Hamilton-Jacobi PDE for the
halved conformal exponent can be approximated by a quasilinear PDE
near a critical point of the potential $V$ and studies the solutions of
the latter, which provide asymptotic approximants for certain
solutions of the contact Hamilton-Jacobi equation. We found that:
{ \be
F(x,u,p)=\frac{a_1(x,u) x^1 p_1+a_2(x) x^2 p_2-b(x,u)}{s_2(x)^3}+\cO(||x||_0^2)~~,
\ee}
\!where $a_i$ and $b$ are homogeneous polynomial functions of degree
six in $x_1$ and $x_2$ (whose coefficients depend on $u$) given by:
{ \be
a_i(x,u)=\lambda_i s_2(x)\left[t_i(x)+6 V(c) e^{2u} s_2(x) s_3(x)\right]~,
\ee}
with:
{\beqa
&& t_1(x)= \lambda_1 \lambda_2^2(\lambda_1-\lambda_2) x_2^2 [s_2(x)-3 \lambda_2 s_1(x)]~~,\nn\\
&& t_2(x)= \lambda_2 \lambda_1^2(\lambda_2-\lambda_1) x_1^2 [s_2(x)-3 \lambda_1 s_1(x)]~~,\nn\\
&&b(x,u)=-\lambda_1^3\lambda_2^3(\lambda_1-\lambda_2)^2 x_1^2 x_2^2 s_1(x)+3 V(c) e^{2u} s_2(x) s_3(x)^2~~,\nn\\
&& s_k(x)\eqdef \lambda_1^k x_1^2+\lambda_2^k x_2^2~~.
\eeqa}

\noindent{\bf Proposition 2.}
The contact Hamilton-Jacobi equation \eqref{HJ} is approximated to
first order in $||x||_0$ by the following quasilinear first order PDE:
\ben
\label{quasilinear}
a_1(x,\phi) x^1 \pd_1 \phi+a_2(x,\phi) x^2 \pd_2 \phi=b(x,\phi)~~.
\een

\noindent {\bf Proposition 3.} With respect to the principal values of
$\Hess(V)(c)$, the general solutions of the linearized Hamilton-Jacobi equation for the
halved conformal factor are as follows:
\begin{itemize}
\item When $\lambda_1\neq \lambda_2$, then the general smooth solution of the linearized equation is:
{\ben
\label{gensol}
\phi(r,\theta)=\phi_0(\theta)+Q_0\left(\frac{\lambda_2-\lambda_1}{\lambda_1\lambda_2} \log r+\frac{1}{\lambda_1}\log|\cos\theta|-\frac{1}{\lambda_2}\log|\sin\theta|\right)~,
\een}
\!where:
{ \be
\label{partsol}
\phi_0(\theta)=\frac{1}{4} \log(\lambda_1^2 \cos ^2\theta+\lambda_2^2 \sin ^2\theta)-
  \frac{1}{2} \frac{\lambda_2 \log |\cos\theta|-\lambda_1\log |\sin\theta|}{\lambda_2-\lambda_1}
\ee}
\!and $Q_0$ is an arbitrary smooth function of a single variable.

\item When $\lambda_1=\lambda_2:=\lambda$, then the linearized equation reduces to:
{\be
\label{lindeg}
x^i\pd_i \phi=\frac{1}{2}~~,
\ee}
\!whose general solution is: 
{\ben
\label{gensoldeg}
\phi(r,\theta)=\frac{1}{2} \log r+\phi_0(\theta)~~,~~\phi_0\in \cC^\infty(\rS^1) ~\mathrm{an~arbitrary~smooth~function}.
\een}
\end{itemize}

\section{Conclusions}

This is a summary of part of the results in \cite{axion1}, where we
investigated the consistency conditions underlying slow-roll
rapid-turn (SRRT) inflation in two-field cosmological models, focusing
in particular on the strong SRRT regime. In this context, the strict
form of the strong consistency condition can be formulated as a
geometric partial differential equation (PDE) that constrains the
interplay between the scalar field metric and the scalar potential
defined on the target manifold. This equation encodes the requirement
that the background trajectory in field space supports sustained
inflationary evolution with a large turning rate while preserving
approximate slow-roll behavior.

When supplemented with suitable boundary or regularity conditions, this
geometric PDE can determine either the scalar field metric
in terms of a prescribed potential or, conversely, the potential
compatible with a given target-space geometry. In this way, the
equation acts as a {\em selection criterion} for “fiducial” models that
realize consistent strong SRRT inflation. Such models form a
distinguished subclass of multifield inflationary models,
characterized by their dynamical stability and by geometric
compatibility between curvature, potential gradients, and the
inflationary turning motion.

When the scalar potential $V$ is specified, the analysis of the strong
SRRT equation can be simplified by fixing the conformal class of the
field-space metric, thereby reducing the coordinate-change freedom in
the geometric sector and facilitating the PDE’s resolution. The
resulting equation then constrains the conformal factor of the metric.

We analyzed this geometric PDE using standard techniques from
quasilinear PDE theory, with particular attention to its behavior
near non-degenerate critical points of the potential, which
correspond to stationary configurations of the scalar fields. By
applying a quasilinearization procedure in a neighborhood of such
critical points, we derived asymptotic expansions which describe the
local structure of admissible solutions. This analysis provides
natural boundary conditions and regularity constraints that may
correspond to local inflationary ``attractors'' in the strong SRRT
regime.

Our approach \cite{axion1} builds upon and extends the framework proposed by
Lazaroiu and Anguelova \cite{cons, consproc}, who identified 
the strong SRRT consistency
condition as a powerful tool for classifying viable two-field
inflationary models. By interpreting this condition as a geometric
PDE, we further clarified its mathematical structure and its role as a
unifying constraint linking kinematical properties of inflationary
trajectories with the underlying field-space geometry and scalar
potential. These results contribute to a more systematic understanding
of the geometric foundations of multifield inflation and open the way
for constructing explicit families of consistent models that can be
tested through future cosmological observations.

The present work suggests numerous new questions and directions for
further research. In particular, one can perform a similar analysis
for the weak SRRT equation. Furthermore, one could write specialized
code to compute efficiently solutions of the contact Hamilton-Jacobi
equation for general Riemann surfaces and could investigate the
problem of existence and uniqueness of globally-defined viscosity
solutions with prescribed asymptotics. Also, one could try to use the
theory of integrable systems to find which fiducial models are integrable.

\section*{Acknowledgments}
\noindent This work was supported by the national grant PN 23210101/2023 and is partly based upon work from COST Action COSMIC WISPers CA21106, supported by COST (European Cooperation in Science and Technology).

\end{document}